\documentclass[12pt]{article}
\usepackage[reqno]{amsmath}
\usepackage{bbm}
\usepackage{epsfig}
\usepackage{array}
\usepackage{float}
\usepackage{dsfont}
\usepackage{amstext}

\usepackage{a4}

\usepackage{a4wide}

\def\be{\begin{equation}}
\def\ee{\end{equation}}
\def\gs{\mathrel{
   \rlap{\raise 0.511ex \hbox{$>$}}{\lower 0.511ex \hbox{$\sim$}}}}
\def\ls{\mathrel{
   \rlap{\raise 0.511ex \hbox{$<$}}{\lower 0.511ex \hbox{$\sim$}}}}

\newcommand{\ba}{\begin{array}{c}}
\newcommand{\baz}{\begin{array}{cc}}
\newcommand{\bad}{\begin{array}{ccc}}
\newcommand{\bea}{\begin{equation} \begin{array}{c}}
\newcommand{\eea}{ \end{array} \end{equation}}
\newcommand{\ea}{\end{array}}
\newcommand{\D}{\displaystyle}
\newcommand{\dms}{\mbox{$\Delta m^2_{\odot}$}}
\newcommand{\dma}{\mbox{$\Delta m^2_{\rm A}$}}
\newcommand{\meff}{\mbox{$\langle m \rangle$}}

\begin{document}

\title{\vspace{-2cm}
\hfill {\small TUM--HEP--598/05}\\
\vspace{-0.3cm} 
\hfill {\small hep--ph/0507312} 
\vskip 0.2cm
\bf 
Broken $\mu$--$\tau$ Symmetry and Leptonic $CP$ Violation 
}
\author{
R.~N.~Mohapatra$^{a,b}$\thanks{email: \tt rmohapat@physics.umd.edu }~\mbox{ 
}~~and~~
W.~Rodejohann$^b$\thanks{email: \tt werner$\_$rodejohann@ph.tum.de} 
\\\\
{\normalsize \it $^a$Department of Physics and Center for String and Particle 
Theory,}\\
{\normalsize \it University of Maryland, College Park, MD-20742, USA}\\ \\ 
{\normalsize \it $^b$Physik--Department, Technische Universit\"at M\"unchen,}\\
{\normalsize \it  James--Franck--Strasse, D--85748 Garching, Germany}
}
\date{}
\maketitle
\thispagestyle{empty}
\vspace{-0.8cm}
\begin{abstract}
\noindent We propose that leptonic $CP$ violation 
arises from a breaking of the $\mu$--$\tau$ exchange symmetry in the 
neutrino mass matrix which in turn is indicated by 
the near maximal atmospheric neutrino mixing 
and the near zero $\theta_{13}$. We find that for the case of a 
normal hierarchy, present data already restricts the way $CP$ violation may 
appear in the neutrino mass matrix and there is an interesting 
correlation between the mixing angle 
$\theta_{13}$, the solar mixing angle $\theta_{12}$ and 
the Dirac $CP$ phase. 
In the inverted hierarchy only $\theta_{13}$ and the Dirac phase 
are linked. 
We also discuss the impact 
of this kind of $CP$ violation on the deviation of the atmospheric mixing 
from its maximal value. Moreover, if corrections to $\mu$--$\tau$ symmetry 
arise from the charged lepton sector, 
where $\mu$--$\tau$ symmetry is known to be broken 
anyway, we find interesting connections
between the $CP$--even and --odd terms as well.  
Our predictions are testable in the proposed 
long baseline neutrino experiments.
\end{abstract}

\newpage

\section{\label{sec:intro}Introduction}
Recent discoveries in neutrino physics have opened up an interesting 
window into physics beyond the standard model. Aside from the very fact 
that the neutrinos have mass, there are two surprising sets of results 
(not in line with expectations based on our knowledge of the quark sector) 
that have provided new opportunities for the field: they are 
(i) the extreme smallness of the overall value of the neutrino masses 
and (ii) the dramatically different mixing pattern from quarks. The first 
one can be interpreted as an indication that 
perhaps a new symmetry such as $B-L$ may be appearing at high scales 
so that one can use the seesaw mechanism using the right--handed neutrinos 
to resolve the puzzle of small masses. The mixing pattern on the other hand
poses a completely different and much more challenging problem. It is 
expected that solving this problem will reveal new symmetries 
for leptons and perhaps as well for quarks. 

A fundamental theory
can of course determine the structure of both the charged lepton and the
neutrino mass matrices and
therefore will lead to predictions about lepton mixings. However, in the
absence of
such a theory, one wants to adopt a model independent approach and look
for symmetries
that may explain the mixing pattern which in turn can throw light on the 
nature of the fundamental theory for quarks and leptons. 

One such symmetry has been inspired by the near maximal value of 
$\theta_{23}$ and the near zero value of $\theta_{13}$. 
These angles appear in the usual definition of the PMNS matrix 
\begin{equation} \label{eq:Upara}
 U= U^\dagger_{\ell} \, U_\nu  = \left(
 \begin{array}{ccc}
 c_{12} \, c_{13} & s_{12}\, c_{13} & s_{13}\, e^{-i \delta}\\
 -c_{23}\, s_{12}-s_{23}\, s_{13}\, c_{12}\, e^{i \delta} &
 c_{23}\, c_{12}-s_{23}\, s_{13}\, s_{12}\, e^{i \delta} & s_{23}\, c_{13}\\
 s_{23}\, s_{12}-\, c_{23}\, s_{13}\, c_{12}\, e^{i \delta} &
 -s_{23}\, c_{12}-c_{23}\, s_{13}\, s_{12}\, e^{i \delta} & c_{23}\, c_{13}
 \end{array}
 \right) ~, 
\end{equation}
where $U_\nu$ diagonalizes the neutrino mass matrix via 
$U_\nu^T \, {\cal M}_\nu \, U = {\cal M}_\nu^{\rm diag}$ and 
$U_\ell$ is associated with 
the diagonalization of the charged leptons. 
If one works in a basis where charged leptons are mass 
eigenstates, then $U^\dagger_{\ell} = \mathbbm{1}$ and 
a simple way to understand $\theta_{23} = \pi/4$ and $\theta_{13} = 0$ 
is to postulate that the neutrinos obey a $\mu$--$\tau$ interchange 
symmetry \cite{mutau,mass,mutau_cp,mutau2,rabimutau}. 
Even though this procedure 
is a 
basis dependent one, 
the hope is that any symmetries for leptons revealed in this
basis are true or approximate symmetries of Nature itself\footnote{One 
might think that since the muon and tau lepton masses are so different, 
such a symmetry cannot be present in Nature. While this could be true, it 
is worth remembering that in the context of supersymmetric models, the 
origin of the up--type fermion masses (i.e., $u,c,t$ quark masses and the 
neutrino masses) and those of the down type fermion masses ($d,s,b$ 
quarks and $e,\mu$ and $\tau$ leptons) arise from different Higgs 
doublets, the so--called $H_u$ and $H_d$ fields. Therefore, one could 
easily imagine a situation where $H_u$ and the quarks are all singlets 
under the $\mu$--$\tau$ exchange symmetry whereas $H_d$, the leptons 
and the right--handed neutrinos are not. In a situation like this, the 
neutrinos could respect $\mu$--$\tau$ symmetry whereas the charged leptons 
need not.}.

A convenient parameterization of the neutrino mass matrix in this basis 
that obeys exact $\mu$--$\tau$ symmetry is given for the case of normal 
hierarchy by \cite{rabimutau}:
\begin{eqnarray} \label{eq:mnu0}
{\cal M}_\nu~=~\frac{\sqrt{\Delta
m^2_{\rm A}}}{2}\left(\begin{array}{ccc}d \, \epsilon^n 
& a \, \epsilon &a \, \epsilon\\ a \, \epsilon & 1 + \epsilon & -1 \\
a \, \epsilon & -1 & 1 + \epsilon\end{array}\right)~, 
\end{eqnarray}
where $n \geq 1$ and $\epsilon \ll 1$. 
An immediate prediction of this mass matrix is that $\theta_{23}=\pi/4$, 
$\theta_{13}=0$ and furthermore $\epsilon~\sim \sqrt{\Delta
m^2_\odot/\Delta m^2_{\rm A}}$.
Note that this matrix conserves $CP$ symmetry. This suggests several ways 
for introducing $CP$ violation in the leptonic sector.

$CP$ violation could either be introduced while maintaining $\mu$--$\tau$ 
symmetry or by breaking it. In the first case, the most general form is
\begin{eqnarray}
{\cal M}_\nu~=~\frac{\sqrt{\Delta
m^2_{\rm A}}}{2}\left(\begin{array}{ccc}d \, \epsilon^n 
& a \, \epsilon \, e^{i\alpha} 
& a \, \epsilon \, e^{i\alpha}
\\ 
a \, \epsilon \, e^{i\alpha} & 
1 + \epsilon & -e^{i\gamma} \\
a \, \epsilon \, e^{i\alpha} & -e^{i\gamma} & 1 + \epsilon\end{array}\right)~.
\end{eqnarray}
While this matrix still leads to maximal atmospheric neutrino mixing 
and vanishing $U_{e3}$, 
it destroys another nice feature, i.e., the smallness of $\sqrt{\Delta
m^2_\odot/\Delta m^2_{\rm A}}$, which is not any more directly 
connected to the small parameter $\epsilon$. 
In fact, the small solar mass squared difference and the 
large solar neutrino mixing require that 
$\gamma$ be very small, i.e., $\gamma \leq \sqrt{\Delta
m^2_\odot/\Delta m^2_{\rm A}}$, so that the 2--3 sub--determinant 
of the mass matrix is small. 
Therefore, in what follows, we set 
$\gamma=0$. This case is essentially the same as the 
$CP$ conserving case since the new phase simply changes the 
Majorana phases in the PMNS matrix and does  
not manifest itself in oscillation phenomena. The same conclusion 
holds if a phase is put in the $ee$ entry. 

On the other hand, if $CP$ violation arises because of  
$\mu$--$\tau$ symmetry breaking, a priori one could imagine putting the $CP$ 
phase in different places in the matrix. 
If we put the phase in the $\mu\mu$ or  $\tau\tau$ entry, then it has to be 
close to zero in order to make the lower 2--3 sub--determinant small. 
Hence, the only position where a phase breaking the 
$\mu$--$\tau$ symmetry could be put is either the $e\mu$ entry or the 
$e\tau$ entry. Both are equivalent and we therefore discuss the case where 
only the $e\tau$ entry of the neutrino mass matrix Eq.\ (\ref{eq:mnu0}) 
is complex \cite{nishiura}. 

The situation for the case of inverted hierarchy is somewhat more 
complicated and is discussed later on in the text.

In this paper, we pursue the 
consequences of this hypothesis for both the normal and inverted hierarchy 
and show that it has several
 interesting consequences, linking for instance the size of 
$\theta_{13}$ and $\theta_{23} - \pi/4$ with the magnitude of leptonic $CP$ 
violation and the value of the solar neutrino mixing angle. These 
correlations depend on whether the neutrino mass hierarchy is 
normal or inverted. 

A different way to break $\mu$--$\tau$ symmetry is when contributions 
from the charged lepton sector perturb the PMNS matrix. One could  
imagine that the neutrino mass matrix is completely $\mu$--$\tau$ symmetric 
and corrections stem entirely from the charged lepton sector, which is known 
to break $\mu$--$\tau$ symmetry anyway. We 
also analyze this case and show there can also be interesting correlations 
between the observables. Unlike the previous case, the results for this 
case are independent of the mass hierarchy of the neutrinos.

\section{\label{sec:NH}Normal hierarchy}
In this section, we discuss the case when the charged lepton mass matrix 
is diagonal and a leptonic $CP$ violating phase is present in the neutrino 
mass matrix in a way that it breaks $\mu$--$\tau$ symmetry.
This leads to very interesting phenomenology. Consider the following 
matrix\footnote{Breaking the symmetry by a complex entry in the 
$e\mu$ instead of the $e\tau$ element will give basically 
the same results, with the only change that $\alpha \rightarrow - \alpha$.}: 
\be \label{eq:mnu_NH}
{\cal M}_\nu = \frac{m_0}{2} 
\left(
\bad 
d \, \epsilon^2 & a \, \epsilon & a \, \epsilon \, e^{i \alpha} \\[0.2cm]
a \, \epsilon & 1 + \epsilon & -1 \\[0.2cm]
a \, \epsilon \, e^{i\alpha} & -1 & 1 + \epsilon 
\ea
\right)~, 
\ee
where $\epsilon \ll 1$ and $a,d$ are real parameters of order 
one\footnote{This particular form of the mass matrix is motivated by 
approximate leptonic symmetries of the type $U(1)_{L_e}$ times 
the $\mu$--$\tau$ exchange symmetry, 
where $\epsilon$ denotes the strength of $U(1)_{L_e}$ 
breaking. This can keep the $\mu$--$\tau$ sector to have entries of 
order one, the $e\mu$ and $e\tau$ elements 
of order of $\epsilon$ and the $ee$ entry of order $\epsilon^2$.}.
It is clear from the discussion above that in the limit 
of $\alpha=0$ exact $\mu$--$\tau$ symmetry is recovered and 
the values $\theta_{23}=\pi/4$ and $U_{e3}=0$ are predicted. 
It is straightforward to obtain the phenomenology of this matrix: 
first, the effective mass governing neutrinoless double beta decay is 
very much suppressed (for a recent overview of 
the situation regarding the effective mass, see, e.g., \cite{fut_data}). 
Turning to $CP$ violation in neutrino oscillations, all such observables are 
described in terms of one rephasing invariant, namely \cite{branco}: 
\bea \label{eq:jcp0}
J_{CP} = \frac{1}{8} \, \sin 2 \theta_{12}\, \sin 2 \theta_{23}\, 
\sin 2 \theta_{13}\, \cos\theta_{13}\, \sin\delta = 
\frac{\D {\rm Im} \left\{ h_{12} \, h_{23} \, h_{31} \right\} }
{\D \Delta m^2_{21} \, \Delta m^2_{31} \, \Delta m^2_{32}~}~, \\[0.3cm]
\mbox{ where } h = {\cal M}_\nu^\dagger\, {\cal M}_\nu~. 
\eea
For the matrix in Eq.\ (\ref{eq:mnu_NH}) we have that 
\be
{\rm Im} \left\{ h_{12} \, h_{23} \, h_{31} \right\} 
\simeq - \frac{m_0^6}{16} \, 
a \, \epsilon^3 \, \sin \alpha + {\cal O}(\epsilon^4)~.
\ee
As it should, this expression (as well as the more lengthy exact one) vanishes 
for $\alpha = 0$.  Of course, one cannot identify $\delta$ 
in the parametrization of the PMNS matrix (\ref{eq:Upara}) with $\alpha$, 
even though $J_{CP}$ is proportional to $\sin \alpha $. 
This can be 
understood as follows: in our model, 
\be \label{eq:res1}
|U_{e3}| \simeq \frac{\D a}{\D 2} \, \epsilon \, 
\sqrt{1 - \cos \alpha} ~~\mbox{ and } ~~
\sin \delta \simeq - \cos \frac{\alpha}{2}~.
\ee
As a result, $\delta$ formally does not vanish for $\alpha = 
0$.  There is no inconsistency in this result, because the invariant 
$J_{CP}$ vanishes if $\alpha = 0$. 
We see from Eq.\ (\ref{eq:res1}) that $\alpha = 0$ corresponds to 
$U_{e3}=0$, which leaves the Dirac phase undefined. What is interesting is 
that there is an inverse correlation between $U_{e3}$ and the Dirac phase 
$\delta$.

\begin{figure}[b]\vspace{-2.5cm}
\hspace{-2cm}
\epsfig{file=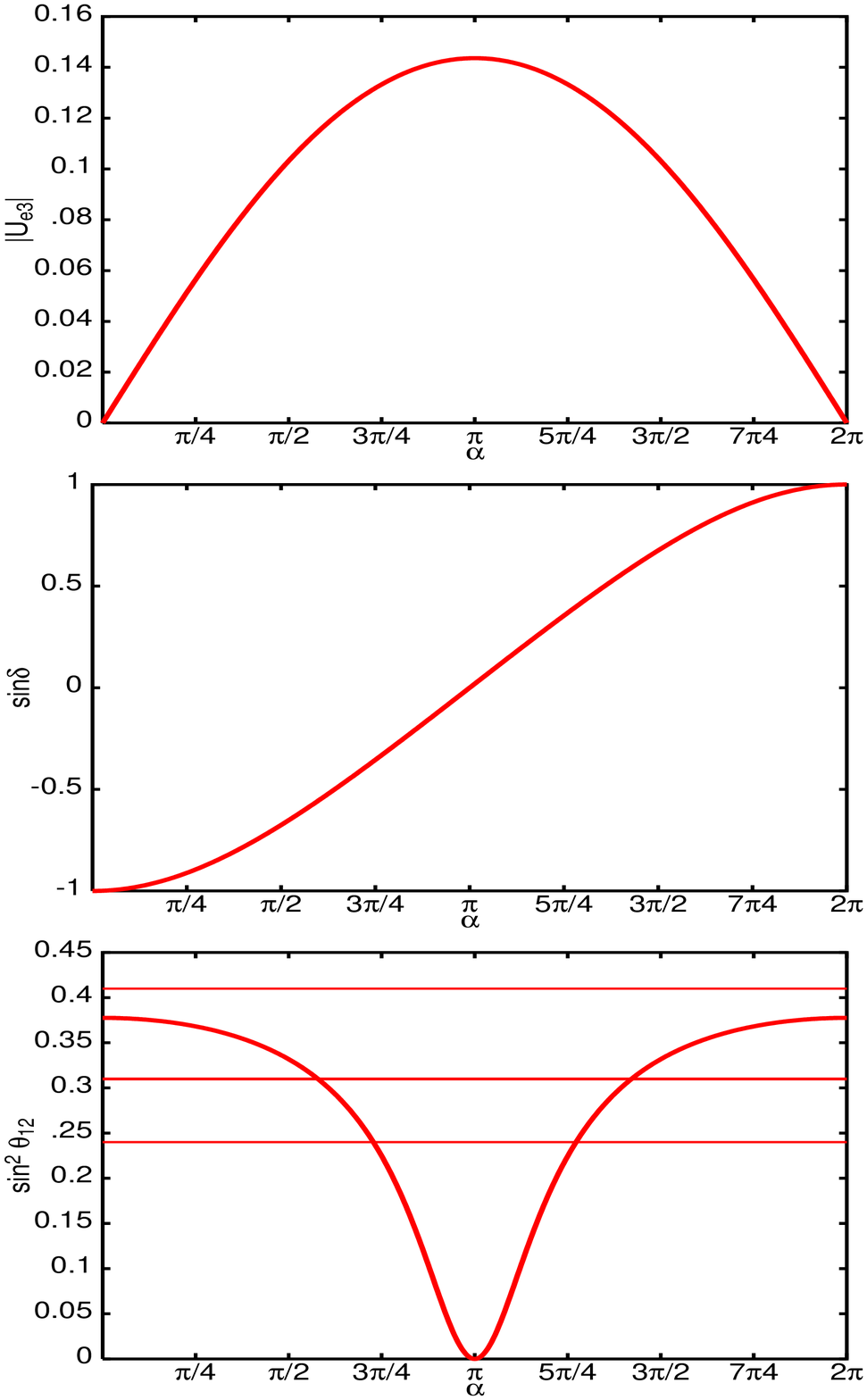,width=19cm,height=24cm}
\caption{\label{fig:NH1}Behavior of $|U_{e3}|$, $\sin \delta$ and 
$\sin^2 \theta_{12}$ as a function of the $\mu$--$\tau$ symmetry 
breaking phase $\alpha$, as it follows from 
Eq.\ (\ref{eq:mnu_NH}). For this particular 
example we chose $a = 1.4$ and $\epsilon = 0.16$. We also 
indicated the current best--fit value and 
$3\sigma$ range of $\sin^2 \theta_{12}$ from \cite{SC_new}.}
\end{figure}

Solar neutrino mixing is governed by  
\be
\tan 2 \theta_{12} \simeq 2 \, a \, \sqrt{1 + \cos \alpha} 
~,  
\ee
which for $\alpha = 0$ agrees with the result from \cite{rabimutau}. 
The value of $\alpha = \pi$, which would mean maximal 
$U_{e3}$ and no $CP$ violation, is not allowed because it would result 
in $\theta_{12} = 0$. 
In Figure \ref{fig:NH1} we show for a particular choice of the parameters 
$a, b, \epsilon$, the behavior of $|U_{e3}|$, $\sin \delta$ and 
$\sin^2 \theta_{12}$ as a function of $\alpha$. 
Though maximal $CP$ violation (i.e., $\sin \delta = \pm 1$) occurs only 
when $U_{e3}=0$, and maximal $|U_{e3}|$ occurs only when $J_{CP}=0$, 
we can have large values of $\sin \delta$ for sizable 
and testable values of $|U_{e3}|$: for instance 
$\sin \delta \simeq 0.5$ with  $|U_{e3}| \simeq 0.05$. 
Instead of scanning the whole parameters space of the model, we note here 
that the following ratio of 
observables depends only on the $\mu$--$\tau$ symmetry 
breaking parameters and not on the order one parameters $a$ or $d$:
\be
\frac{|U_{e3}|}{\tan 2 \theta_{12}} \simeq \frac{\epsilon}{4} 
\, \tan \frac{\alpha}{2} \simeq \frac{\epsilon}{4} \cot \delta~.
\ee

Atmospheric neutrino mixing stays very close to maximal 
and is of course maximal for a vanishing phase:  
\be
\theta_{23} - \frac{\pi}{4} \simeq -\frac{a }{2} \, \epsilon^2 \, 
\sin \alpha/2~.  
\ee
Hence, the deviation from $\sin^2 \theta_{23} = 1/2$ is of order 
$\epsilon^2$, so that next generation experiments \cite{chef} will not 
be able to measure any non--maximality. 
The required precision can only 
be achieved by far--future projects like neutrino 
factories or $\beta$--beams.

Finally, the mass squared differences are given by 
\bea
\dma \simeq m_0^2 \, (1 + \epsilon) ~, \\[0.2cm]
\dms \simeq \frac{\D m_0^2}{\D 4} \, 
\sqrt{1 + 4 \, a^2  \, (1 + \cos \alpha)} \, \epsilon^2~,
\eea
where we set $d \, \epsilon^2$ to zero to keep tractable expressions. 
We see that as in the case of $\mu$--$\tau$ breaking with real parameters,  
the ratio of solar and atmospheric $\Delta m^2$ is of 
order $\epsilon^2$, which is also the order of $|U_{e3}|^2$ \cite{rabimutau}. 
The presence of a non--trivial phase can however slightly 
spoil this simple behavior. 
The results for the oscillation parameters were obtained by the usual 
perturbative method, in which first the 23--block of Eq.\ (\ref{eq:mnu_NH}) 
is diagonalized, then the 13-- and after that the 12--block. Details of 
this standard procedure can be found, e.g., in Ref.\ \cite{king}. 

A different $\mu$--$\tau$ breaking is possible when 
the $e\mu$ and $e\tau$ entries are complex conjugates \cite{mutau_cp}: 
\be \label{eq:mnu_2}
{\cal M}_\nu = \frac{m_0}{2} 
\left(
\bad 
d \, \epsilon^2 & a \, \epsilon \, e^{-i \alpha} 
& a \, \epsilon \, e^{i \alpha} \\[0.2cm]
 a \, \epsilon \, e^{-i \alpha} & 1 + \epsilon & -1 \\[0.2cm]
a \, \epsilon \, e^{i \alpha} & -1 & 1 + \epsilon 
\ea
\right)~.
\ee
The predictions of this matrix are 
$\theta_{23} - \frac{\pi}{4} \simeq a^2 \, \epsilon^2 \, 
\cos \alpha \, \sin \alpha 
$, maximal $CP$ violation (i.e., $\delta = \pm \pi/2$) and 
$|U_{e3}| \simeq a/\sqrt{2} \, \sin \alpha \, \epsilon$. 
Moreover, we find that  
$\tan 2 \theta_{12} \simeq 2\sqrt{2} \, a \, \cos \alpha$ and 
$\dms \simeq m_0^2 /4 \, \sqrt{1 + 8 a^2 \, \cos^2 \alpha}$.  
Though the phase $\alpha$ does not affect the 
size of the low energy phase $\delta$, it is again crucial for the magnitude 
of $|U_{e3}|$ and for obtaining large solar neutrino mixing: 
a maximal $U_{e3}$ is obtained for $\alpha = \pi/2$, which would mean 
that $\theta_{12}=0$. 

Since in seesaw models any symmetry of the neutrino mass matrix is 
supposed to manifest only at the seesaw scale, it is of interest to ask 
what radiative corrections would do to this result. 
It is well known that radiative corrections are small in case of a normal 
hierarchy. For instance, 
using the approximative formulae from Ref.\ \cite{chef2}, we can estimate 
the $\beta$--function for the Dirac phase. 
Neglecting the smallest mass state $m_1$ and using that 
$|U_{e3}| \simeq \sqrt{\Delta
m^2_\odot/\Delta m^2_{\rm A}}$, we estimate 
that for the MSSM $\dot{\delta} \sim 10^{-6} \, (1 + \tan^2 \beta)$. 
In case of just the Standard Model (SM), we have $\dot{\delta} \sim 10^{-6}$. 
To put it another way, let us denote the 
parameter responsible for radiative corrections with $\epsilon_1$, 
where $\epsilon_1 = 
c \frac{m_\tau^2}{16 \pi^2 \, v^2} \ln \frac{M_X}{m_Z}$, with 
$c$ given by 3/2 in the SM and by $-(1 + \tan^2 \beta)$ 
in the MSSM. 
As long as $\epsilon_1 \ll 1$ (or $\tan \beta \ls 50$) 
and $\epsilon_1 \ll \epsilon$, the predictions given above, 
in particular the one of maximal $CP$ violation, 
remain unaffected.

The mass matrix from Eq.\ (\ref{eq:mnu_NH}) could be modified by 
making the $ee$ entry proportional to $d \, \epsilon$, a 
form which however could not be achieved by simple models 
based on $U(1)$ charges. In this case the effective mass in neutrinoless 
double beta decay would be 
larger by a factor of $1/\epsilon \sim \sqrt{\dma/\dms}$ and the 
expression for solar neutrino mixing would be 
$\tan 2 \theta_{12} \simeq 2 \, a \, \sqrt{1 + \cos \alpha}/(1 - d)$. 
The expressions for $J_{CP}$ and $\dms$ would be more complicated, all other 
observables are however as above.

\section{\label{sec:IH}Inverted hierarchy}
Now we break $\mu$--$\tau$ symmetry for the case of inverted hierarchy. 
In this case, the phase that breaks the symmetry could be located in 
various places. Here we consider the simplest case where
the mass matrix is a special case of the broken flavor symmetry 
$L_e - L_\mu - L_\tau$ \cite{lelmlt1,FPR} and reads 
\be \label{eq:mnu_IH}
{\cal M}_\nu = m_0 
\left(
\bad 
d \, \epsilon & a  & a \, e^{i \alpha} \\[0.2cm]
a & b \, \epsilon & f \, \epsilon  \\[0.2cm]
a \, e^{i \alpha}  & f \, \epsilon  & b \, \epsilon 
\ea
\right)~, 
\ee
where $\epsilon \ll 1$ and $a, b, d, f$ are real and of order one. 
Since obviously $m_0 \simeq \sqrt{\dma}$ and the lower 
limit on the effective mass 
$\meff = m_0 \, d \, \epsilon$ is given by \cite{fut_data}
$\meff \gs \cos 2 \theta_{12} \, \sqrt{\dma} \simeq 0.4 \, \sqrt{\dma} $, 
we see that $\epsilon$ should not be too small. 

The Jarlskog invariant is proportional to 
\be
{\rm Im} \left\{ h_{12} \, h_{23} \, h_{31} \right\} \simeq 
2 \, \epsilon^2  \, a^4  \, b \, (f + (b + d) \, \cos \alpha ) \sin \alpha 
+ {\cal O }(\epsilon^3)~.
\ee
This and the 
exact expression 
vanish for $\alpha = 0$ as in the case of normal hierarchy. 
Diagonalizing Eq.\ (\ref{eq:mnu_IH}) gives (the procedure is described, e.g., 
again in Ref.\ \cite{king}) 
\bea \label{eq:res1_IH}
\theta_{23} - \frac{\pi}{4} \sim \epsilon^2 \, \sin \alpha~,~ 
|U_{e3}| \simeq \frac{\D b}{\D \sqrt{2} \, a} \, \epsilon \, 
\sin \alpha ~~\mbox{ and } ~~
\sin \delta \simeq \cos \alpha~. 
\eea
Again, as in the case of normal hierarchy, $\delta$ does formally not 
vanish for 
$\alpha =0$. However, in this limit $J_{CP}$ vanishes, 
which can be understood because $U_{e3}=0$ for $\alpha =0$. 
Solar neutrino mixing is typically rather large, because it is in first order 
inversely proportional to $\epsilon$. A short approximate formula 
is possible when the entries in the lower $\mu\tau$ block of 
Eq.\ (\ref{eq:mnu_IH}) are of order $\epsilon^2$, in which case 
$|U_{e3}| \simeq \frac{b}{\sqrt{2} \, a} \, \epsilon^2 \, 
\sin \alpha$, together with 
$\theta_{23} - \frac{\pi}{4} \simeq \frac{bd}{\sqrt{2}a} \, 
\epsilon^3 \, \sin \alpha $ and 
$\tan 2 \theta_{12}  \simeq \frac{2\sqrt{2} \, a }{d \, \epsilon}$. 
For both possibilities, however, 
the strong dependence of $\theta_{12}$ on the phase, which we encountered 
in case of a normal hierarchy, is lost. 
The large value of $\theta_{12}$ reflects the well--known fact that 
the texture in Eq.\ (\ref{eq:mnu_IH}) in the limit of $\epsilon = 0$ 
produces maximal solar neutrino mixing. Sizable 
contributions from the charged lepton 
sector are then required to reach accordance with the data \cite{FPR,QLC}. 
For $\theta_{12}^\nu = \pi/4$, small $|U_{e3}|^\nu \neq 0$ 
and for the most natural case of 
hierarchical charged lepton mixing the result is 
\be
\sin^2 \theta_{12} \simeq \frac{1}{2} - \frac{1}{\sqrt{2}} 
\sin \theta_{12}^\ell ~\mbox{ and } 
|U_{e3}| \simeq \left| 
|U_{e3}|^\nu - \frac{\sin \theta_{12}^\ell}{\sqrt{2}} \, 
c_\alpha \right| 
~,
\ee
plus higher order terms. The deviation from maximal solar neutrino 
mixing is connected to the magnitude of $U_{e3}$, where however 
cancellations can occur.  
We will give a more detailed analysis on charged lepton contributions 
below. 

We can again analyze the case of the $e\mu$ element being the 
complex conjugate of the $e\tau$ element. 
Maximal $CP$ violation in the sense of $\delta = \pm \pi/2$ is predicted, 
as well as $|U_{e3}| \simeq \epsilon \, \sqrt{2} \, b/a \, \cos \alpha \, 
\sin \alpha$ and $\theta_{23} - \pi/4 \propto \epsilon^3 \, \sin \alpha$. 
Again,  
$\tan 2 \theta_{12}$ is large because at first order it is 
inversely proportional to $\epsilon$. 

As in the case of a normal hierarchy, 
effects of radiative corrections are rather small. Here the reason is that 
the pseudo--Dirac structure of the mass matrix (\ref{eq:mnu_IH}) 
implies that the two heaviest mass states 
have basically opposite $CP$ parities, which largely suppresses the 
running. Neglecting the smallest mass state and 
using $|U_{e3}| \simeq \sqrt{\Delta
m^2_\odot/\Delta m^2_{\rm A}}$, one can estimate that 
$\dot{\delta} \sim 10^{-5}~(1 + \tan^2 \beta)\, \Delta m^2_{\rm A}/\Delta
m^2_\odot \, \sin \Delta \phi$, 
where $|\Delta \phi| \simeq \pi$ is the difference of the 
Majorana phases of the two leading mass states.

\section{\label{sec:CL}Contributions from the Charged Lepton Sector}
We saw above that for the inverted hierarchy case and 
if we are too close to the $L_e-L_\mu-L_\tau$ symmetry limit, one 
might require contributions from the charged lepton sector to fit 
observations.  
Here we analyze this possibility in more detail in order 
to investigate whether there are correlations between the $CP$--even 
and --odd observables for this case as well.
Recall that the charged leptons strongly break $\mu$--$\tau$ 
symmetry anyway and therefore the 
case when the neutrino sector is $\mu$--$\tau$ 
symmetric 
and corrections stem from the charged 
lepton sector is surely appealing.

The basic idea is that the matrix $U_\ell$ associated with the diagonalization 
of the charged leptons corrects a given neutrino mixing scheme.  
This could be the bimaximal \cite{FPR,QLC,devbimax} or the tribimaximal 
\cite{devtri,PR} mixing scenario or, as in our case, the 
$\mu$--$\tau$ symmetric scenario \cite{xing}. In the last case, the 
neutrino mixing 
matrix (ignoring the Majorana phases) can be written as: 
\be
\tilde{U}_\nu = 
\left( 
\bad 
\cos \theta_{12}^\nu  & \sin \theta_{12}^\nu & 0 \\[0.3cm]
\frac{\D - \sin \theta_{12}^\nu }{\D \sqrt{2}} 
& \frac{\D \cos \theta_{12}^\nu }{\D \sqrt{2}} 
& \frac{\D 1}{\D \sqrt{2}} \\[0.3cm]
\frac{\D \sin \theta_{12}^\nu }{\D \sqrt{2}} 
& \frac{\D - \cos \theta_{12}^\nu }{\D \sqrt{2}} 
& \frac{\D 1}{\D \sqrt{2}}
\ea 
\right) ~.
\ee
The angle $\theta_{12}$ is a priori undefined, for instance, it was quite 
large for the matrix (\ref{eq:mnu_IH}) and defined by 
$\tan 2 \theta_{12} =  2 \sqrt{2}a/(1 - d \epsilon)$ for Eq.\ 
(\ref{eq:mnu_NH}) and $CP$ conservation \cite{rabimutau}. 
Note that the following discussion is independent on the neutrino mass 
spectrum. 

\begin{figure}[b]\vspace{-2.5cm}
\hspace{-2cm}
\epsfig{file=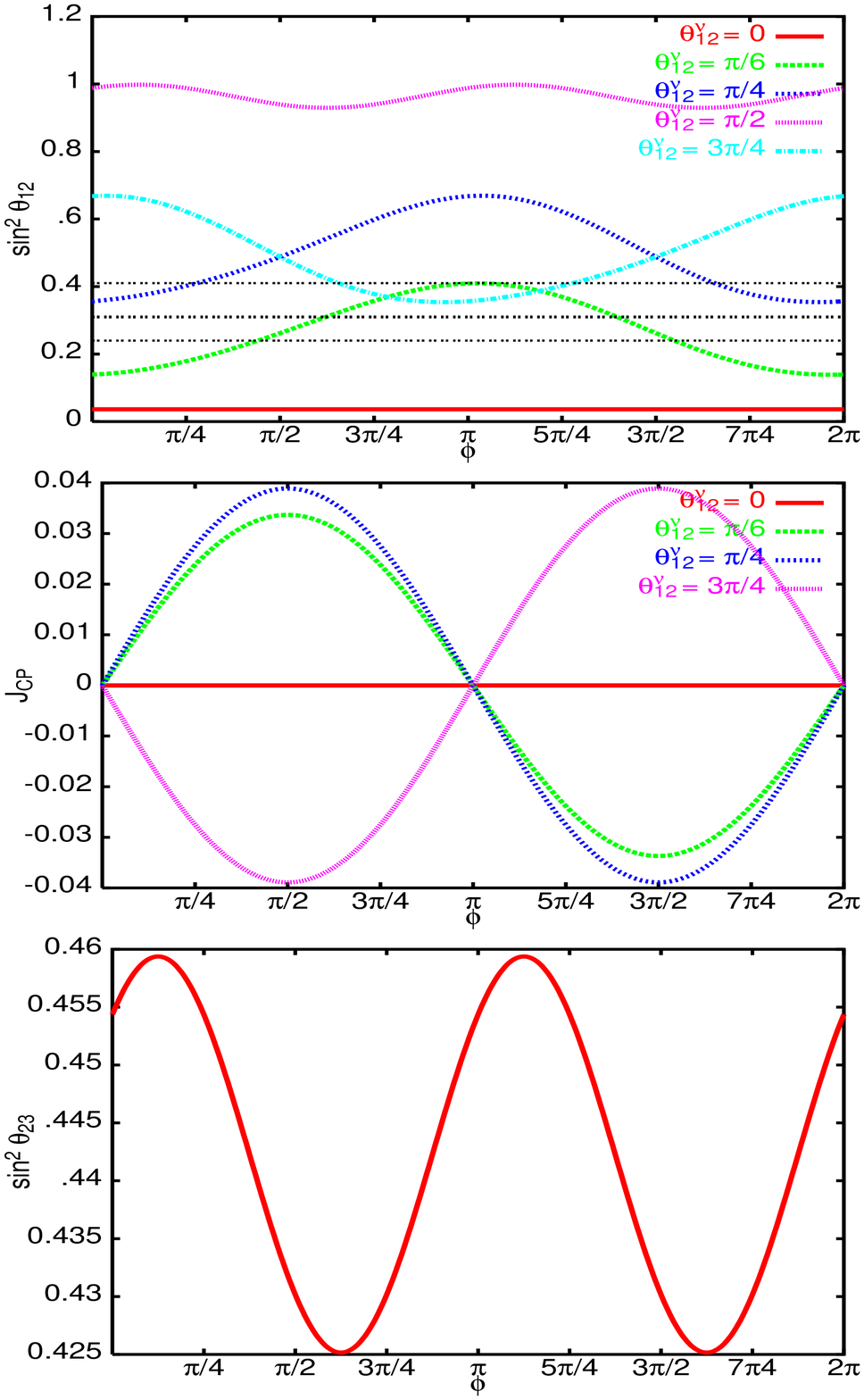,width=19cm,height=24cm}
\caption{\label{fig:CL}Behavior of $\sin^2 \theta_{12}$ and $J_{CP}$ 
as a function of the phase $\phi$, as it follows from 
Eq.\ (\ref{eq:ser_obs_2}). In this example $|U_{e3}| \simeq 0.16$ and 
$\psi = 60^0$, $B = 0.8$ and $\omega = 90^0$.}
\end{figure}

As has been shown in Ref.\ \cite{FPR}, one can in general express the 
PMNS matrix as 
\be\label{eq:parapmns}
U =  
\tilde{U}_{\ell}^\dagger \, P_\nu \, \tilde{U}_\nu \, Q_\nu ~.
\ee
It consists of two diagonal phase matrices 
$P_\nu$ = diag($1,e^{i \phi}, e^{i \omega}$) and 
$Q_\nu$ = diag($1,e^{i \rho}, e^{i \sigma}$), as well as two 
``CKM--like'' matrices $\tilde{U}_\ell$ and $\tilde{U}_\nu$ which contain 
three mixing angles and one phase each, and are parametrized in analogy to 
Eq.\ (\ref{eq:Upara}).  Out of the six phases present in 
Eq.\ (\ref{eq:parapmns}), five stem from the neutrino 
sector. We denote the phase in $\tilde{U}_{\ell}$ with $\psi$. 
The six phases will in general 
contribute in a complicated manner to the three observable ones. 
 In case of one angle in $U_\nu$ being zero (a result of $\mu$--$\tau$ 
symmetry), the 
phase in $\tilde{U}_\nu$ is unphysical. 
Note that the 2 phases in $Q_\nu$ do not appear in 
observables describing neutrino oscillations \cite{FPR}. 
Let us assume that the matrix $U_\ell$ contains only small angles, whose 
sines we denote by $\lambda_{ij} \equiv \sin \theta_{ij}^\ell$, with 
$ij=12,13,23$. 
Keeping only the first order terms in the $\lambda_{ij}$, we 
get\footnote{Setting in these equations $\theta_{12}^\nu$ to $\pi/4$ 
(to $\sin^{-1} \sqrt{1/3}$) reproduces the formulae 
from \cite{FPR} (\cite{PR}).}  
\bea \label{eq:ser_obs_1}
\sin^2 \theta_{12} \simeq \sin^2 \theta_{12}^\nu  - 
\frac{1}{\sqrt{2}} \, c_\phi \, \sin 2 \theta_{12}^\nu \, \lambda_{12} 
+ \frac{1}{\sqrt{2}} \, c_{\omega - \psi} \, 
\sin 2 \theta_{12}^\nu \, \lambda_{13} ~,\\[0.3cm]
|U_{e3}| \simeq \frac{1}{\sqrt{2}} \,  
\left| \lambda_{12} + \lambda_{13} \, e^{i(\omega - \psi - \phi)}  \right|
~,\\[0.3cm]
\sin^2 \theta_{23} \simeq \frac{1}{2} - \lambda_{23} \, c_{\omega - \phi}~, 
\eea 
where we introduced the notations 
$c_\phi=\cos\phi$, $c_{\omega - \psi} = \cos(\omega - \psi)$ and so on.
The parameter 
$\lambda_{23}$ does in first order not appear in $\sin^2 \theta_{12}$ and 
$|U_{e3}|$, just as $\lambda_{12,13}$ does not in $\sin^2 \theta_{23}$. 
We see from the expressions that --- unless there are cancellations --- 
$|U_{e3}|$ is lifted from its value 
zero, and that the same small parameters $\lambda_{12,13}$ appear in leading 
order also in the deviation from the original value of 
$\theta_{12}$.  
Leptonic $CP$ violation is described by 
\be
J_{CP} \simeq \frac{1}{4 \sqrt{2}} \, \sin 2 \theta_{12}^\nu \, 
\left( 
 \lambda_{12} \, s_\phi + \lambda_{13} \, s_{\omega - \psi} 
\right)~.
\ee
Consider now the case of a strong hierarchy in the charged lepton mixing, 
say, a ``CKM--like'' structure in the form of 
$\lambda_{12}=\lambda$, 
$\lambda_{23}=A \, \lambda^2$ and $\lambda_{13}=B \, \lambda^3$ 
with $A$, $B$ real and of order one. 
Then we have 
\bea \label{eq:ser_obs_2}
\sin^2 \theta_{12} \simeq \sin^2 \theta_{12}^\nu  - 
\frac{1}{\sqrt{2}} \, c_\phi \, \sin 2 \theta_{12}^\nu \, \lambda  
+ \frac{1}{4} \, 
\left( 
2 \, (c_{2 \phi} + s_{2 \phi}) \, \sin^2 \theta_{12}^\nu 
+ 3 \, \cos 2 \theta_{12}^\nu 
\right)\, \lambda^2 ~,\\[0.3cm]
|U_{e3}| \simeq \frac{1}{\sqrt{2}} \, \lambda ~,\mbox{ where } 
J_{CP} \simeq \frac{\lambda}{4 \sqrt{2}} \, \sin 2 \theta_{12}^\nu 
\, s_\phi ~,\\[0.3cm]
\sin^2 \theta_{23} \simeq \frac{1}{2} + \frac{1}{4} 
\left( 
c_{2 \phi} + s_{2 \phi} - 4 B c_{\omega - \phi} - 2
\right) \, \lambda^2~, 
\eea 
plus terms of ${\cal O}(\lambda^3)$. We see that the leading 
contribution to $\sin^2 \theta_{12}$ depends on the same phase $\phi$ to which 
leptonic $CP$ violation is proportional. Since it follows that at leading 
order 
\be
\sin^2 \theta_{12} \simeq \sin^2 \theta_{12}^\nu  - \sin 2 \theta_{12}^\nu 
\, |U_{e3}| \, c_\phi ~, 
\ee
there are some constraints on the possible values of $\phi$, $|U_{e3}|$ 
and $\theta_{12}^\nu$. Fixing $\lambda = 0.22$ 
(which can be motivated by ideas such as Quark--Lepton--Complementarity 
\cite{QLC}), we give in Fig.\ \ref{fig:CL} the implied 
correlations between the parameters. 
We see that $|U_{e3}| \simeq \lambda/\sqrt{2} \simeq 0.16$ and that 
atmospheric neutrino mixing can deviate sizably from maximal. The 
parameter $\theta_{12}^\nu$, which fixes the relative size of the parameters  
in a $\mu$--$\tau$ symmetric mass matrix, 
has to lie somewhere between $\pi/6$ and $3\pi/4$. As can be seen, 
large $CP$ violation is possible in these cases.

\section{\label{sec:sum}Summary}
In summary, we have considered the possibility that leptonic $CP$ 
violation owes its origin to the 
breaking of a $\mu$--$\tau$ symmetry, which in turn is motivated by 
the near maximal $\theta_{23}$ and the near zero $\theta_{13}$. 
One consequence of this hypothesis is that interesting correlations 
between the size 
of $U_{e3}$, the magnitude of $CP$ violation and even the solar 
neutrino mixing angle occur if the neutrino mass hierarchy is normal. 
Radiative corrections to these conclusions 
are mild and the 
results are easily testable in future neutrino experiments. For the case 
of inverted hierarchy there is no correlation between the solar neutrino 
mixing and the $CP$ phase. 
We furthermore studied the possibility that the 
$\mu$--$\tau$ breaking originates
in the charged lepton sector with the neutrino sector being $\mu$--$\tau$ 
symmetric and presented its consequences.
Interestingly, also in this case there can be significant correlations 
between $CP$--even and --odd observables.

\vspace{0.5cm}
\begin{center}
{\bf Acknowledgments}
\end{center}
The work of R.N.M.~was supported by the National Science Foundation 
grant no.\ Phy--0354401 and the 
Alexander von Humboldt Foundation (the Humboldt Research Award). 
The work of W.R.~was supported by the
the ``Deutsche Forschungsgemeinschaft'' in the 
``Sonderforschungsbereich 375 f\"ur Astroteilchenphysik'' 
and under project number RO--2516/3--1 (W.R.). R.N.M.\ would like to 
thank M.\ Lindner for discussions and hospitality at TUM during the time 
this work was done.

\pagestyle{empty}

\end{document}